\documentclass{pasj00}
\usepackage{booktabs}
\usepackage{color}

\begin{document}
\SetRunningHead{K. Maeda and H. Shibahashi}{Pulsations of Pre-White Dwarfs with Hydrogen-dominated Atmospheres}
\Received{2014/03/31}
\Accepted{2014/05/18}
\Published{}

\title{Pulsations of Pre-White Dwarfs with Hydrogen-dominated Atmospheres}

\author{Kazuhiro \textsc{Maeda} and Hiromoto \textsc{Shibahashi}
}
\affil{Department of Astronomy, The University of Tokyo, Bunkyo-ku, Tokyo 113-0033}
\email{3txqufj@gmail.com {\rm and} 
shibahashi@astron.s.u-tokyo.ac.jp}


\maketitle

\begin{abstract}
We carried out a fully non-adiabatic analysis for nonradial oscillations of pre-white dwarfs evolved from the post-Asymptotic Giant Branch (AGB) with hydrogen-dominated envelopes. It is shown that nuclear reactions in the hydrogen burning-shell excite low-degree g-modes in the period range of about 40-200\,s for the pre-white dwarf models with $T_{\rm eff}=$ 40\,000\,K - 300\,000\,K. 
It is also shown that the amount of hydrogen {has} a significant influence on the instability domain of such pre-white dwarfs in the Hertzsprung-Russel (H-R) diagram. 
Thus, the thickness of hydrogen-dominated envelopes may be well constrained by observing the presence of the g-mode oscillations. 
\end{abstract}

\section{Introduction}
White dwarfs are at the end stage of stars born initially with mass less than about eight times the Sun. The vast majority of stars, over 95\% of the stars in the Galaxy, end their lives as white dwarfs. 
Thus, it is worth investigating white dwarfs at various stages in their cooling sequences to extend our understandings of cosmochronology {and stellar evolution theories}. 

It is thought that the main channel of stellar evolution to the white dwarf sequence is from the post-AGB stars, going through the violent mass loss and planetary nebula phase. It is also thought that about three quarters of post-AGB stars enter the white dwarf domain with hydrogen-rich atmospheres. The remaining one quarter of post-AGB stars undergo a very late helium-shell flash that pushes them back to the AGB again. Such an evolution is called the `born-again' process. This produces a violent mixing that leads to burn a substantial amount of residual hydrogen away. These stars then exhibit atmospheres made of mixture dominated by He, C, and O, and gravitational segregation eventually makes these stars have helium-rich atmospheres.
In addition to the evolution channel from the AGB phase, a small fractions {($\sim$\,2\%)} of white dwarfs 
come out of the Extreme Horizontal Branch (EHB) phase. 
For more details about evolution of white dwarfs, readers are recommended to consult recent review papers such as \citet{Althaus2010}, \citet{Fontaine2012}, and \citet{Fontaine2013}.

White dwarfs are classified according to their spectroscopic features.
About 85\% of white dwarfs reveal absorption lines of only hydrogen \citep{Eisenstein2006}, indicating that they have pure hydrogen atmospheres at the photospheric level.
This purity is thought to be a consequence of strong gravity of these stars. Chemical elements are segregated by gravity according to atomic weight, and if a large amount of hydrogen remains in the atmosphere of stars at the pre-white dwarf stage, 
it quickly levitates to form pure hydrogen-atmosphere white dwarfs, which are called the DA stars. 
The bulk of the rest shows only helium lines, indicating a lack of hydrogen in their atmospheres, and such helium atmosphere white dwarfs are classified as DO and DB stars, depending on whether the absorption lines are ionized helium ions or neutral helium. Other classes of white dwarfs are classified according to the presence of polluting elements, and they are known as DQ, DZ, and DC stars.

Pre-white dwarfs, stars just before becoming white dwarfs as a result of consumption of nuclear energy generation sources, still have double burning shells; a hydrogen-burning shell and a helium-burning shell.
It is well known that interactions between these burning shells induce a thermal instability. 
The resultant thermal flashes and consequent structural changes are taken into account {in} theoretical calculation of evolution of stars. 
Shell burning can also be pulsationally unstable \citep{Kippenhahn1983}. Indeed, nonradial g-modes driven by the $\varepsilon$-mechanism were found for models of hydrogen-deficient post-planetary nebulae nuclei \citep{Kawaler1986}. However, the effect of such pulsational instability on evolution of stars is uncertain. 
\citet{Corsico2009} report that short periodicities observed in a hydrogen-deficient star VV\,47 could be due to the $\varepsilon$-mechanism powered by an active helium-burning shell. 
\citet{Corsico2009} extensively carried out stability analyses of hydrogen deficient hot pre-white dwarfs and  revealed the presence of pulsationally unstable short-period g-modes destabilized by the helium-burning shell through {the $\varepsilon$-mechanism.
These} are located in a well-defined theoretical instability strip in the $\log T_{\rm eff}$-$\log g$ diagram. They also showed that the $\varepsilon$-driven g-modes have enough time to reach detectably large amplitudes.

We should be reminded here that the majority of white dwarfs are DA stars with pure hydrogen atmospheres rather than the helium-rich, hydrogen-deficient atmospheres. 
We should then investigate the pulsational stability of pre-white dwarfs with hydrogen-dominated atmospheres.
\citet{Kawaler1988} carried out a stability analysis of models of hydrogen shell-burning planetary nebula nuclei, and found that instability occurs indeed.
\citet{Kawaler1988} concluded, however, that the absence of observed pulsation (affirmed by \cite{HN1987}) suggests that the central stars do not retain enough hydrogen to support nuclear burning following ejection of the planetary {nebulae}. 

Later, 
the instability due to hydrogen burning was found in {post-EHB} stars \citep{Charpinet1997}. 
{\citet{Handler1998} searched for periodic light variations for nine such stars, including five of the stars recommended by \citet{Charpinet1997}, but no pulsations with amplitude larger than $\sim 1$\,mmag were detected. 
However, spectroscopic estimates for these nine stars have been revisited recently, and all these objects are now considered as post-AGB stars and not as post-EHB stars \citep{Gianninas2010}.} We anticipate that {highly precise observation with better instruments or bigger telescopes may provide} an opportunity to actually detect a luminosity variation that could not be detected decades ago. Also the theoretical investigation of pulsational stability of pre-white dwarfs with hydrogen-rich atmospheres so far made should be extended significantly more to get a definitive conclusion.  
Hence, we would like to stress the importance of re-examining pulsational instabilities due to the $\varepsilon$-mechanism in pre-DA white dwarfs in terms of both theoretical study and observation. 

In this paper, we revisit the $\varepsilon$-mechanism in pre-DA white dwarfs that have substantial nuclear burning in their hydrogen shells. 
In such pre-white dwarfs, the amount of hydrogen is supposed to be crucial in discussing the instabilities due to the $\varepsilon$-mechanism. 
However, in the studies of \citet{Kawaler1988} and \citet{Charpinet1997},  the relation between the envelope thickness and the excitation in burning shell was not fully investigated. 
To fulfill this demand, we systematically investigate pulsational stability of pre-DA white dwarfs, by adopting the Modules for Experiments in Stellar Astrophysics (MESA) \citep{Paxton2011, Paxton2013} to compute pre-white dwarf models by following evolution from the Zero-Age Main-Sequence (ZAMS). 
By tuning parameters (the initial mass and the mass loss parameter), we obtain pre-white dwarf models with different amounts of hydrogen. 
We then focus on the relation between the envelope thickness and the instability strip.

\section{Equilibrium Models}
\subsection{Evolutionary models}
We construct pre-white dwarf models by following evolution from the ZAMS, using the MESA stellar evolution code. 
The solar abundance reported by \citet{Asplund2009} is adopted as the initial chemical composition. 
We target at carbon-oxygen core white dwarfs in order to study the most common type of white dwarfs.
Hence helium- or oxygen-neon core white dwarfs are excluded from the models to be studied in this paper.

Since the basic aim of this paper is to investigate the relation between the hydrogen-envelope thickness and the pulsational instability, 
models with different amounts of hydrogen need to be constructed. 
Toward this purpose, we tuned the initial mass, $M_{\rm ini}$, and the mass loss parameter, $\eta$, to obtain distinct values of the total hydrogen mass in the envelope.
Other parameters were set to be common in all the calculations. 
The initial mass was chosen between $1.5\,M_\odot$ and $6.0\,M_\odot$ at proper intervals.  
The mass loss schemes adopted in this work are that of \citet{Reimers1975} for red giants and \citet{Blocker1995} for AGB stars. 
Their formulae are respectively described as 
\begin{eqnarray} 
	\dot{M}_{\rm R}(\eta) &=&4\times 10^{-13}  \eta\frac{LR}{M}~~[M_\odot/{\rm yr}]~ , \\
	\dot{M}_{\rm B}(\eta) &=& 4.83\times 10^{-9} \frac{L^{2.7}}{M^{2.1}}  \dot{M}_{\rm R}(\eta), 
\end{eqnarray}
where $L$, $R$, and $M$ are the luminosity, the radius and the mass of the star in solar units. 
Thus, by tuning the parameter set of $(M_{\rm ini}, \eta)$, we have obtained models with different amounts of hydrogen $M_{\rm H}$ in the envelopes. 
Some sets of parameters failed to provide a convergence, and we sometimes found the so-called born-again process. 
After the model sequences undergo the born-again, some of them have their hydrogen envelopes completely depleted and thereby maintain helium-atmospheres, whereas some still have a substantial amount of hydrogen remaining in the envelopes. 
Such calculations were discarded in order just to treat the simplest evolutions. 
The models {took into account diffusion}, by which the gravitational segregation process was computed. 

The evolution calculations were done with a standard number of mesh points from the ZAMS to a certain point in the post-AGB phase. 
After that point, the number of mesh points was increased (about 1.5 times), 
and also the atmospheric structure was then added to the point where the optical depth $\tau$ is $10^{-4}$. 

\begin{table}  
\caption{Model sequences used in this work. Each model is calculated from the ZAMS and the models at $\log L/L_\odot \sim -2.0$ are listed in this table to compare them in the same scheme. 
The initial mass, the total mass, the radius, and the total hydrogen mass are shown in solar units.
}
\begin{center}
\begin{tabular}{cccccc}
\toprule
label & $M_{\rm ini}$ & $\eta$ & $M$ & $R$ & $M_{\rm H}$ \\ 
\midrule
1 & 1.5 & 2.9 & 0.504 & 0.0151 & 1.2 $\times 10^{-4}$ \\
2 & 1.5 & 2.7 & 0.517 & 0.0148 & 1.1 $\times 10^{-4}$ \\
3 & 1.5 & 2.1 & 0.532 & 0.0144 & 9.7 $\times 10^{-5}$ \\
4 & 3.0 & 22 & 0.545 & 0.0142 & 8.6 $\times 10^{-5}$ \\
5 & 3.0 & 14 & 0.560 & 0.0139 & 7.6 $\times 10^{-5}$ \\
6 & 3.0 & 5.0 & 0.577 & 0.0136 & 6.6 $\times 10^{-5}$ \\
7 & 3.5 & 16 & 0.610 & 0.0130 & 5.0 $\times 10^{-5}$ \\
8 & 3.5 & 3.0 & 0.645 & 0.0124 & 3.7 $\times 10^{-5}$ \\
9 & 3.8 & 4.0 & 0.700 & 0.0116 & 2.3 $\times 10^{-5}$ \\
10 & 4.0 & 2.0 & 0.746 & 0.0110 & 1.6 $\times 10^{-5}$ \\
11 & 4.2 & $3.0\times 10^{-2}$ & 0.801 & 0.0103 & $9.6 \times 10^{-6}$ \\
12 & 4.2 & $1.0\times 10^{-2}$ & 0.816 & 0.0101 & $8.4 \times 10^{-6}$ \\
13 & 4.3 & $3.0\times 10^{-2}$ & 0.840 & 0.0097 & $6.7 \times 10^{-6}$ \\
14 & 4.8 & $3.0\times 10^{-2}$ & 0.868 & 0.0094 & $5.1 \times 10^{-6}$ \\
15 & 6.0 & 0.10 & 0.916 & 0.0088 & $3.1 \times 10^{-6}$
\\
\bottomrule
\end{tabular}
\end{center}
\label{table:models}
\end{table}

\begin{figure}   
\begin{center}
\includegraphics[width=0.75\linewidth,angle=270]{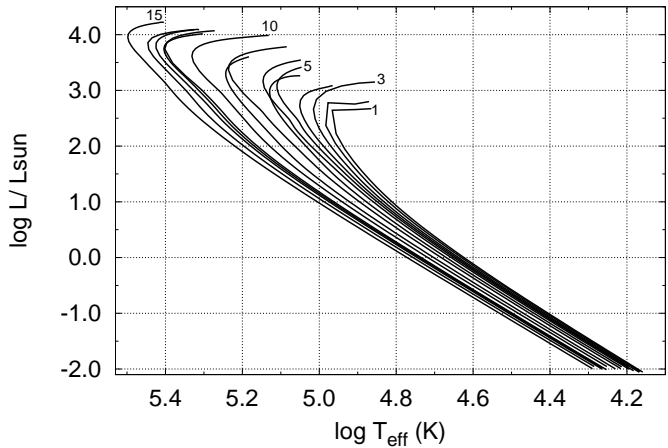}
\end{center}
\caption{Evolutionary tracks of the model sequences used in the present work. 
The label numbers correspond to those listed in table\,\ref{table:models}. }
\label{fig:1}
\end{figure}

Table \ref{table:models} shows the parameters of evolution of models and their final states (at $\log L/L_\odot \sim -2.0$).
As seen from this table, higher initial mass and less mass loss tend to result in a more massive core. 
In addition, we see that the envelope becomes thinner as the total mass increases. 
This is because more massive cores tend to have more active burning-shells in the evolution and there would be less amount of hydrogen and helium in the envelope. 
Their evolutionary tracks are shown in figure\,\ref{fig:1}.
The model sequence 1 is the hydrogen-richest sequence. We regard this sequence as the upper limit case because model sequences having more hydrogen than this model sequence eventually end up undergoing the born-again evolution.
We tried to construct a model with $M_{\rm ini} > 6\,M_\odot$ to obtain a further thinner envelope model.
However, in this mass range, we could not obtain even a single model that converged after a certain number of calculation steps. 
Also, models in this mass range tend to have a rather long computational time required to reach the white dwarf stage. 
Thus, we did not keenly calculate models with an initial mass over $6\,M_\odot$.

\subsection{Structure as pulsating stars}
\begin{figure}[tbp]   
\begin{center}
\includegraphics[width=0.6\linewidth,angle=270]{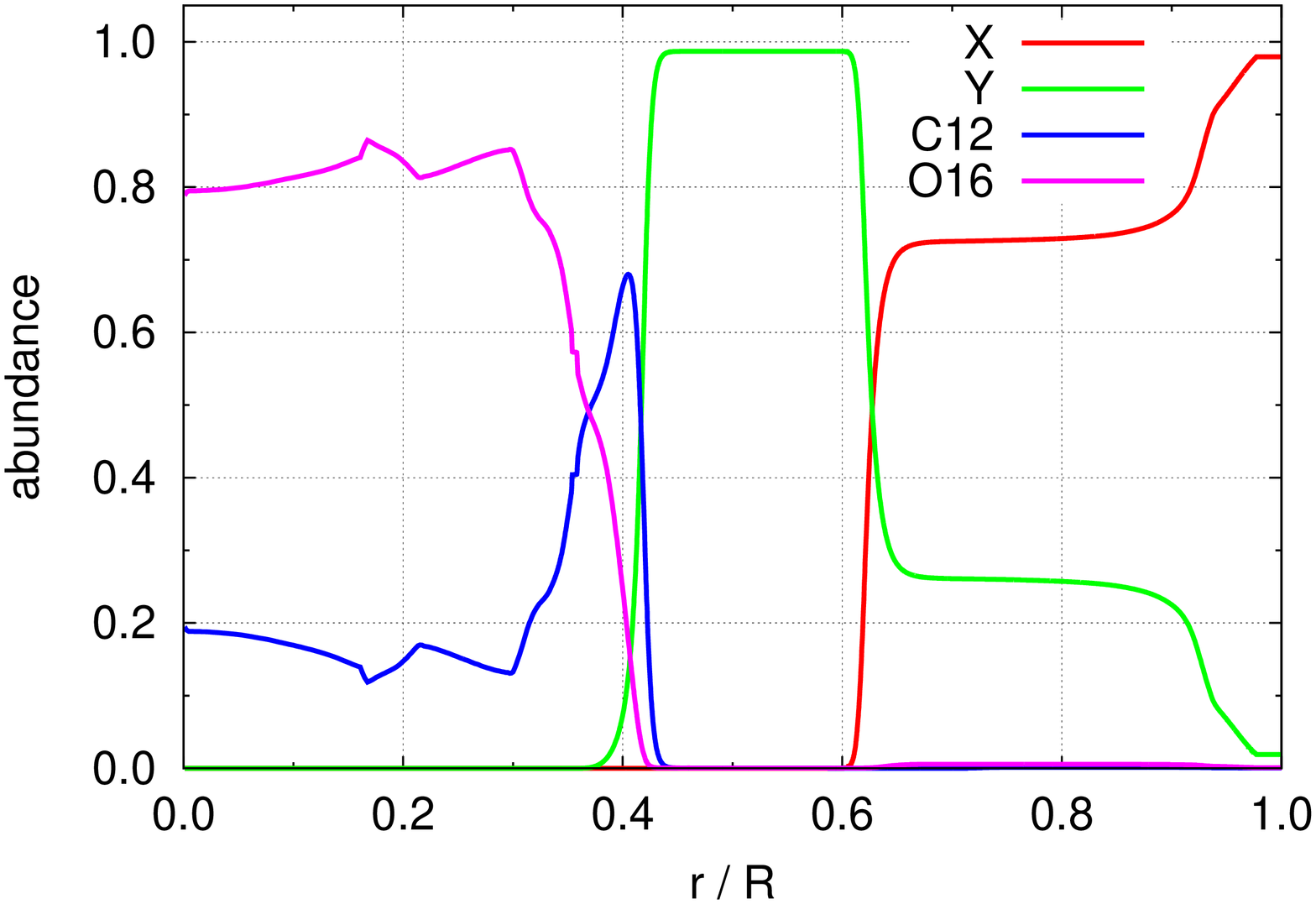}\\
\includegraphics[width=0.6\linewidth,angle=270]{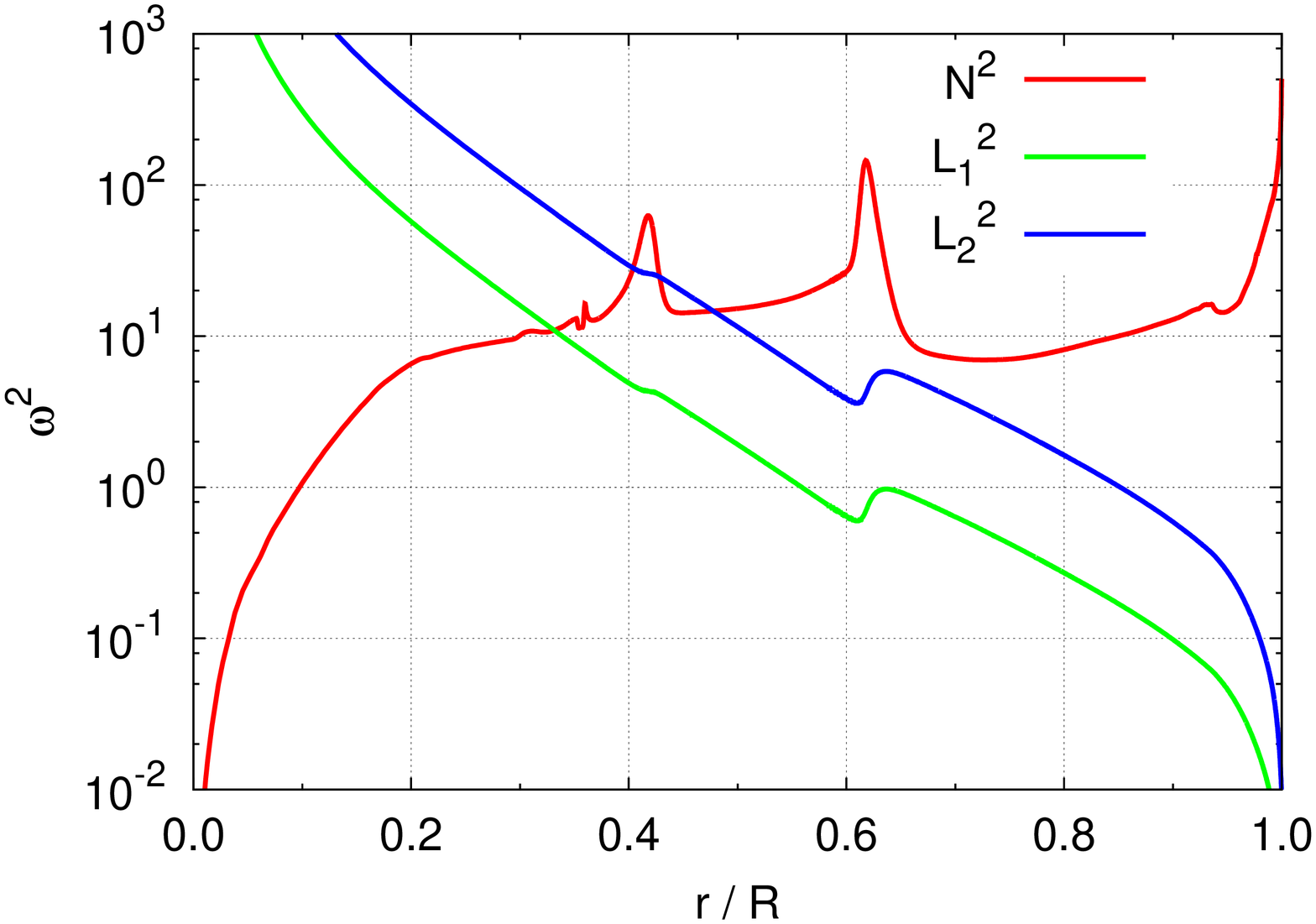}
\end{center}
\caption{Upper panel: The profile of chemical composition in a pre-white dwarf model with $M=0.5323\,M_\odot$, $\log L/L_\odot=1.7336$, and $\log T_{\rm eff}=4.94$. The mass fraction of hydrogen is shown by the red curve, and those of helium, carbon, and oxygen are indicated by the green, blue, and magenta curves. Lower panel: The profiles of the squared Brunt-V\"ais\"al\"a frequency, $N^2$, and the squared Lamb frequency, $L_\ell^2$, for $\ell=1$ and $\ell=2$, in this pre-white dwarf model. The frequencies are measured in units of $(GM/R^3)^{1/2}$.}
\label{fig:2}
\end{figure}

One of the important characteristics of structure of pre-white dwarfs is the presence of double burning-shells. As seen in the upper panel of figure\,\ref{fig:2}, the most outer envelope is composed of a mixture of elements, of which main component is hydrogen. Atomic diffusion causes the helium abundance in the outermost layers to decrease. Hydrogen is consumed as a consequence of hydrogen burning, and the main component is switched to helium in the thin burning-shell. The zone just below the hydrogen-burning shell is mainly composed of helium, and helium is converted to carbon and oxygen in the thin burning-shell at the bottom of this zone. Hence the mean molecular weight changes dramatically at each of the burning-shells, leading sharp peaks of the Brunt-V\"ais\"al\"a frequency. 
It should be remarked that the presence of double burning-shells induces the expansion of the outer zones of each shell, and then the radii of pre-white dwarfs are about twice of the white dwarfs at the later stage. 

The bottom panel of figure\,\ref{fig:2} shows the profile of the squared Brunt-V\"ais\"al\"a frequency in a pre-white dwarf model, along with the squared Lamb frequency, as a function of the fractional radius. The model belongs to the sequence 3, and the mass of the star, the luminosity, and the effective temperature at this stage are
$0.5323\,M_\odot$, $\log L/L_\odot=1.7336$, and $\log T_{\rm eff}=4.94$, respectively.  
Here, the Brunt--V\"ais\"al\"a frequency, $N$, is defined as
\begin{equation}  
	N^2 = {{g}\over{H_p}}
	\left[ \left({{\partial\ln\rho}\over{\partial\ln T}}\right)_p(\nabla-\nabla_{\rm ad}) 
	+\left({{\partial\ln\rho}\over{\partial\ln\mu}}\right)_{p,T}\nabla_\mu
	\right],
\end{equation}
where $H_p := -dr/d\ln p$, $\nabla := d\ln T/d\ln p$, $\nabla_{\rm ad} := (\partial\ln T/\partial \ln p)_S$, and $\nabla_\mu := d\ln\mu/d\ln p$. Here, $g$ denotes the local gravity, $p$ is the pressure, $\rho$ is the density, $T$ is the temperature, and $\mu$ denotes the mean molecular weight. The Lamb frequency, $L_\ell$, is defined as
\begin{equation}  
	L_\ell^2 = {{\ell(\ell+1)c^2}\over{r^2}}.
\end{equation}
Here, $c^2 := \Gamma_1 p/\rho$ denotes the squared sound speed, where $\Gamma_1 := (\partial\ln p/\partial\ln\rho)_S$, and $\ell$ denotes the spherical degree of the mode. 
The steep gradient of chemical composition in the burning-shells appears in this propagation diagram as the peaks of $N^2(r)$ around $r/R \sim 0.42$ and $r/R \sim 0.62$, and a jump in $L_{\ell}^2(r)$ around $r/R\sim 0.62$. There is also a similar, but small jump near $r/R\sim 0.42$.

For a normal mode, the eigenfunction is separated into the temporal part and the spatial part. The former is described as $\exp(i\sigma t)$, and the latter is separated further into the angular part, which is described in terms of spherical harmonic functions, and the radial part. The latter is well written with the WKBJ approximation,
and the wave energy density $E_{\rm W}$ of gravity modes, which propagate in the region where $N^2 > \omega^2$ and $L_\ell^2 > \omega^2$, have amplitudes proportional to 
\begin{equation}  
	E_{\rm W} \propto {{\ell}\over{\omega}} 
	{{N}\over{r}}\sin^2 \left( {{\ell}\over{\omega}}\int {{N}\over{r}}\,dr\right). 
\end{equation}
Here, $\omega^2$ denotes the square of dimensionless frequency $[\,\omega^2 := \sigma^2/(GM/R^3)]$, and $N^2$ and $L_\ell^2$ are measured in the same unit.
This means that the amplitude of a g-mode can be large mainly in a burning-shell, at which the Brunt-V\"ais\"al\"a frequency has a sharp peak, if 
\begin{equation}  
	{{\ell}\over{\omega}} \int {{N}\over{r}} \,dr \simeq (n+1/2)\pi
\end{equation}
in the burning-shell, where $n$ denotes an integer corresponding to the number of nodes there.
This approximation is accurate for high-degree, high-order modes, but it is still useful for a qualitative understanding of general properties of nonradial oscillations. 
For example,
in the case of the hydrogen-burning shell in the model shown in figure\,\ref{fig:2}, the peak value of $N$ is $\sim 10$, the location is $r/R\sim 0.6$, and the thickness is $\sim 0.05\,R$, hence the mode with $\omega\simeq 1$ satisfies the above condition with $n=0$ for $\ell=1$.

Since the radiative dissipation becomes significant as the wavelength becomes short, low-degree and low-order g-modes {should be more easily excited, so should the modes having a large amplitude in the burning-shell}, ---that is, those with no node in the burning-shell. 
One might worry about the fact that the propagation zone of g-modes extends to the deeper region. The displacement due to the oscillation is, however, small in the deep interior, and the same is true for the temperature perturbation, because the amplitude of the displacement is inversely proportional to the square root of the equilibrium density, which is much larger in the deep interior. Hence, the dissipation in the deep interior is not necessarily large to overcome the destabilizing effect in the hydrogen-burning shell.

\section{Instability Analysis}
\subsection{Untable modes vs stable modes}
The white dwarf model sequences shown in table \ref{table:models} were analyzed by a fully non-adiabatic code that was newly developed for the present work.  
The modes were calculated for the degree $\ell=1$\,-\,$15$ and for the radial order $n \leq 20$. 
The instability analysis was applied to each cooling sequence from the planetary nebula nuclei phase to the white dwarf phase ($\log L/L_\odot \sim -2.0$), and a typical model sequence has approximately 70 models analyzed. 

Since the criterion for occurrence of convection is reduced to $N^2 < 0$, the pre-white dwarf model shown in figure \ref{fig:2} is fully radiative. This means the treatment of convection in the non-adiabatic analysis, which is a difficult problem, is avoided in this model. Models in the later evolution stages have convection due to ionized hydrogen, but the $\varepsilon$-mechanism in those models does not work sufficiently enough to cause an instability. Thus, the effect of convection on the non-adiabatic analysis of pulsation is ignorable in the later stages. On the other hand, models in the earlier stages have convection due to ionized heavy elements. However, the ionization zone in such models is located near the photosphere, which is far from the burning-shell. Thus, the influence of convection on the non-adiabatic analysis of pulsation is tiny in the earlier stages. 

\begin{figure}   
\begin{center}
\includegraphics[width=0.7\linewidth,angle=270]{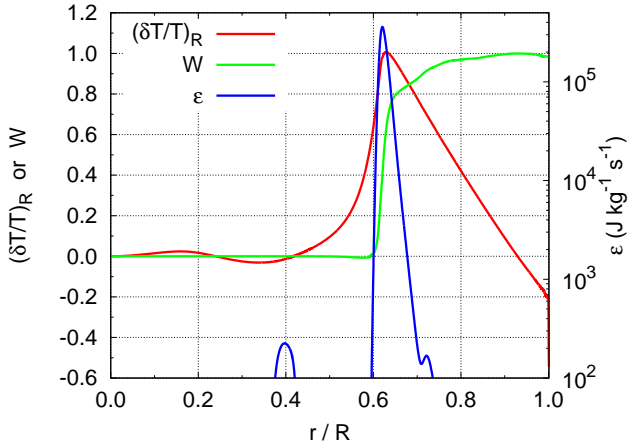}
\end{center}
\caption{The red curve shows the eigenfunction of Lagrangian temperature perturbation, $\delta T/T$, of $\ell=2$ g$_3$-mode ($\omega^2_{\rm R} = 1.39$) of the stellar model shown in figure\,\ref{fig:2}, as a function of the fractional radius. The amplitude has a peak (normalized as unity) in the hydrogen-burning shell, of which nuclear energy generation rate, $\varepsilon$, is plotted with the blue curve. The work integral, $W$,  of the mode is shown with the green curve, which is normalized as unity at its maximum. A sudden increase of $W$ at the hydrogen-burning shell indicates that this mode is excited by the $\varepsilon$-mechanism there.}
\label{fig:3}
\end{figure}
\begin{figure}[!b]    
\begin{center}
\includegraphics[width=0.7\linewidth,angle=270]{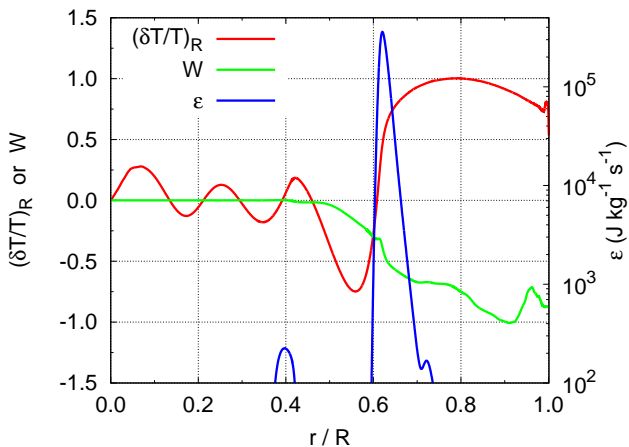}
\end{center}
\caption{The same as figure\,\ref{fig:2}, but for $\ell=1$ g$_7$-mode ($\omega^2_{\rm R} = 0.175$).
The eigenfunction has a node in the hydrogen-burning shell, and reaches its maximum outside the shell. As a consequence, the radiative dissipation overcomes the destabilizing effect of the $\varepsilon$-mechanism, and the mode is not excited.}
\label{fig:4}
\end{figure}

Figure\,\ref{fig:3} shows the eigenfunction of the Lagrangian temperature perturbation of the quadrupole ($\ell=2$) g$_3$-mode of the equilibrium model shown in figure\,\ref{fig:2}. As expected in the previous section, the temperature perturbation of this mode has a large amplitude mainly in the hydrogen-burning shell. As shown by the green curve in figure\,\ref{fig:3}, the work integral of this mode clearly shows that this mode is indeed excited by the $\varepsilon$-mechanism working in the hydrogen-burning shell.  

The excitation mechanism does not efficiently work for g-modes that are not so well trapped in the hydrogen-burning shell. Figure\,\ref{fig:4} shows such an example.
In this case, there is a node in the hydrogen-burning shell and then the amplitude of this mode is not large enough there to induce destabilization of the mode. The radiative dissipation in the outer layers dominates the contribution from the burning-shell.

\begin{figure}   
\begin{center}
\includegraphics[width=0.7\linewidth,angle=270]{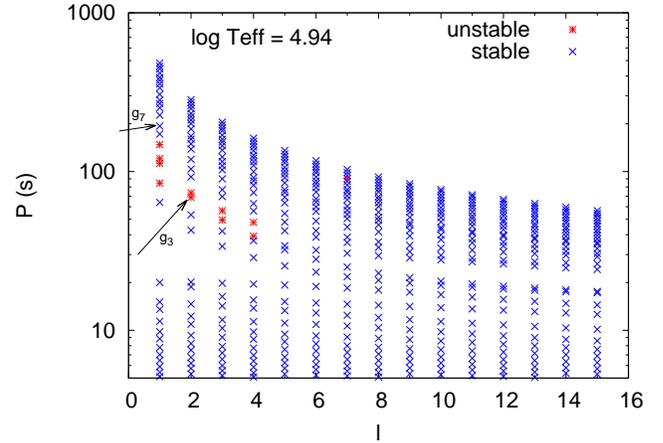}
\end{center}
\caption{Periods of oscillation modes of the model shown in figure\,\ref{fig:2}. The blue ``$\times$'' marks stand for stable modes and the red ``$*$'' marks stand for unstable modes. Only the low-degree, low-order g-modes trapped well in the hydrogen-burning shell are excited, and all the other modes are damped. The two arrows indicate the modes shown in figures \ref{fig:3} and \ref{fig:4}.}
\label{fig:5}
\end{figure}

Figure \ref{fig:5} shows the oscillation periods of the eigenmodes as a function of the spherical degree $\ell$ of the mode. 
We see that it is low-degree, low-order g-modes that are excited  by the $\varepsilon$-mechanism in pre-white dwarf models. 
Low-degree, high-order g-modes and
high-degree g-modes are stabilized by the radiative dissipation in the outer envelope.
The periods of the unstable modes are in the range of $40$\,-\,$200$\,s. 
We have verified this range is also valid for the other model sequences. 

\begin{figure}    
\centering
\includegraphics[scale=0.34,angle=270]{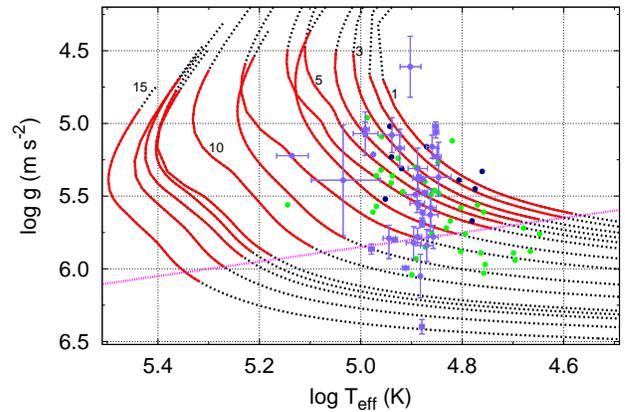} 
\caption{
Evolutionary tracks ($ \log T_{\rm eff}$-$\log g$ diagram) of the sequences in table \ref{table:models}. 
The dotted lines correspond to models analyzed and the numbers shown in the figure correspond to the labels defined in table \ref{table:models}.
The bold red lines correspond to models having at least one unstable mode. 
The straight dotted magenta line may be a good indicator for the cool edge of the instability strip.
{The location of some DA and DAO stars (DA stars exhibiting helium in their optical spectra) are also shown.
Black dots correspond to the stars, of which light variations were searched for by \citet{Handler1998}. 
Their atmospheric parameters are based on \citet{Gianninas2010}.
Error bars are not given in the original paper.
Green dots are the other DAO and hot DA stars listed in tables 2 and 3 of \citet{Gianninas2010}.
Violet dots with error bars are the hot DA stars listed in table 1.3 of \citet{Sion2011}.}
}
\label{fig:6}
\end{figure}

\subsection{Instability along stellar evolution}
{The modes we have found are destabilized solely in the hydrogen-burning shell, and 
contribution of the helium-burning shell is, at best, minor} at least in the modes we have checked. 
Figure \ref{fig:6} is a $\log T_{\rm eff}$-$\log g$ diagram for each model sequence and shows the instability domains. 
We see that models with more hydrogen have an instability domain extended to the lower effective temperature. 
Different sequences have different masses and one might naively suppose the mass of the star itself may have an influence on the instability domain to some extent. 
However, as seen in figure\,\ref{fig:1}, the cooling sequences of white dwarf in the $\log T_{\rm eff}$-$\log L$ plane (the H-R diagram) is confined in a relatively narrow region. 
This means that, at a given effective temperature, the luminosities are not significantly different. 
Thus we conclude that the dependence of total mass of the star on the instability domain is not important since the total amount of the nuclear reaction is not significantly different.  

The envelope thickness greatly affects the instability, due to the reasons explained below. 
The location of the bottom of the hydrogen envelope in terms of radius, not mass coordinate, varies significantly according to the amount of hydrogen. 
While the $\varepsilon$-mechanism works to grow perturbations in the hydrogen-burning shell, which is at the bottom of the envelope, the radiative dissipation works to stabilize them mainly near the surface. 
If the hydrogen-burning shell is located deeper in the interior, the distance to the surface is relatively large.
Then the kinetic energy of pulsations near the surface is not significantly large, since the outer part of the hydrogen envelope is evanescent.
Hence the pulsation has a {lower} stabilization contribution near the surface and is more likely to be unstable. 
In contrast, in case of a thin hydrogen envelope, pulsations still have a substantial amount of kinetic energy near the surface. 
Thus, such pulsations are stabilized near the surface and do not become unstable modes.

Figure \ref{fig:6} also shows that the cool edge of the instability domain in terms of effective temperature greatly varies among the model sequences with different hydrogen amounts. 
In particular, for hydrogen-richer DAs, we expect to find pulsations even at a relatively low effective temperature, such as $\log T_{\rm eff} \sim 4.6~ (T_{\rm eff} \sim 40\,000~{\rm K})$. 
Since the lower limit of $\log T_{\rm eff}$ is thus sensitive to $M_{\rm H}$, 
{it should be} possible to constrain the amount of hydrogen of pre-DA white dwarfs by checking {the presence of pulsational} instabilities. 
For example, at $\log T_{\rm eff} = 5.0$, if a DA pre-white dwarf is a pulsator, then we can constrain the mass of total hydrogen to be higher than about $3.0 \times 10^{-5} M_\odot$. 

\begin{table}   
\caption{List of unstable modes with the shortest growth time in the models in {\bf sequence 3}.
Age is set to be zero for the first pulsating model.
Most of the unstable modes grow fast compared with the evolutionary timescale,
hence they must become finite amplitude oscillations.
However, the growth rates of unstable modes of the model at the end of the instability domain eventually become the same order of magnitude as the evolution timescale of the star.
}
\begin{center}\begin{tabular}{cc|ccrc}
\toprule
age (yrs) & $\log T_{\rm eff}$ (K) & $\ell$ & $n$ & $P$ (s) & $\tau_e$ (yrs) \\ 
\midrule
0 & 5.02 & 1 & 7 & 185.5 & 2.13 $\times 10^3$ \\ 
1.7 $\times 10^4$ & 5.01 & 1 & 5 & 145.6 & 5.04 $\times 10^2$ \\ 
3.7 $\times 10^4$ & 5.00 & 1 & 5 & 141.6 & 6.81 $\times 10^2$ \\ 
5.9 $\times 10^4$ & 4.99 & 1 & 4 & 131.4 & 4.84 $\times 10^2$ \\ 
8.5 $\times 10^4$ & 4.98 & 1 & 4 & 128.0 & 5.08 $\times 10^2$ \\ 
1.5 $\times 10^5$ & 4.95 & 1 & 4 & 122.5 & 7.47 $\times 10^2$ \\ 
1.9 $\times 10^5$ & 4.94 & 1 & 4 & 120.5 & 1.00 $\times 10^3$ \\ 
2.3 $\times 10^5$ & 4.93 & 2 & 3 & 67.6 & 1.20 $\times 10^3$ \\ 
2.9 $\times 10^5$ & 4.91 & 2 & 3 & 66.5 & 1.42 $\times 10^3$ \\ 
4.2 $\times 10^5$ & 4.89 & 2 & 3 & 64.7 & 2.14 $\times 10^3$ \\ 
5.0 $\times 10^5$ & 4.88 & 1 & 3 & 109.4 & 2.60 $\times 10^3$ \\ 
5.9 $\times 10^5$ & 4.86 & 1 & 3 & 108.2 & 3.28 $\times 10^3$ \\ 
8.1 $\times 10^5$ & 4.84 & 1 & 3 & 106.2 & 5.93 $\times 10^3$ \\ 
9.5 $\times 10^5$ & 4.82 & 1 & 3 & 105.5 & 8.76 $\times 10^3$ \\ 
1.1 $\times 10^6$ & 4.80 & 1 & 3 & 105.1 & 1.49 $\times 10^4$ \\ 
1.3 $\times 10^6$ & 4.79 & 2 & 2 & 59.2 & 2.42 $\times 10^4$ \\ 
1.5 $\times 10^6$ & 4.77 & 1 & 2 & 98.4 & 5.31 $\times 10^4$ \\ 
1.8 $\times 10^6$ & 4.75 & 1 & 2 & 99.9 & 1.10 $\times 10^5$ \\ 
2.0 $\times 10^6$ & 4.74 & 3 & 1 & 37.3 & 3.67 $\times 10^5$ \\ 
2.4 $\times 10^6$ & 4.71 & 2 & 1 & 50.3 & 6.74 $\times 10^5$ \\ 
2.7 $\times 10^6$ & 4.69 & 2 & 1 & 51.1 & 1.22 $\times 10^6$ \\ 
3.2 $\times 10^6$ & 4.67 & 1 & 1 & 84.6 & 1.88 $\times 10^6$ \\ 
3.6 $\times 10^6$ & 4.65 & 1 & 1 & 86.5 & 4.28 $\times 10^6$\\  
\bottomrule
\end{tabular}
\end{center}
\label{table:period}
\end{table}

We found, in some models, the so-called strange-modes (see, e.g.,  \cite{Saio1998}, \cite{Godart2014}, and \cite{Sonoi2014} for more details), which  
have so far been found to have an extremely short growth time in theoretical calculations for stars with very high luminosity-to-mass ratio. 
The pre-white dwarfs models analyzed in this paper, of course, have a high luminosity-to-mass ratio, especially around the knee of the evolution tracks.
We have confirmed that, as models evolve, some g-modes suddenly become unstable with short growth times and that they are also powered by the $\varepsilon$-mechanism, but 
we do not discuss the strange-modes in this paper.

\subsection{Growth rates of unstable modes}
Table \ref{table:period} lists the mode having the shortest growth time ($e$-folding time) in each model of {sequence 3}.
Here the growth time is defined as the reciprocal of the imaginary part of the eigenfrequency; $\tau_e := 1/|\sigma_{\rm I}|$, where $\sigma_{\rm I}$ is the imaginary part of the eigenfrequency.
The model sequence 3 should be regarded as a hydrogen-rich model sequence. 
Less hydrogen-rich model sequences cross their instability domains with a relatively short time (table \ref{table:period10}). 
As seen from tables \ref{table:period} and \ref{table:period10}, the {eventual instability strip} would be, to some extent, smaller than that shown in figure\,\ref{fig:6} because there may not be any modes which have enough time to grow near the edges of the instability domain.

\begin{table}[htb]    
\caption{Same as table\,\ref{table:period} but for the model sequence 10. 
Some of these models do not have unstable modes {with growth time} shorter than 
the evolution timescale.}
\begin{center}\begin{tabular}{cc|ccrc}
\toprule
age (yrs) & $\log T_{\rm eff}$ (K) & $\ell$ & $n$ & $P$ (s) & $\tau_e$ (yrs) \\
\midrule
0 & 5.30 & 1 & 9 & 127.3 & 5.75 $\times 10^2$ \\ 
9.4 $\times 10^1$ & 5.33 & 1 & 6 & 94.2 & 1.14 $\times 10^1$ \\ 
2.4 $\times 10^2$ & 5.31 & 1 & 5 & 94.0 & 1.30 $\times 10^1$ \\ 
4.2 $\times 10^2$ & 5.27 & 2 & 5 & 56.5 & 2.63 $\times 10^1$ \\ 
8.5 $\times 10^2$ & 5.26 & 2 & 5 & 56.8 & 3.78 $\times 10^1$ \\ 
1.4 $\times 10^3$ & 5.25 & 2 & 5 & 56.8 & 4.68 $\times 10^1$ \\ 
4.3 $\times 10^3$ & 5.24 & 1 & 5 & 95.9 & 8.47 $\times 10^1$ \\ 
1.3 $\times 10^4$ & 5.21 & 1 & 5 & 93.7 & 2.05 $\times 10^2$ \\ 
2.2 $\times 10^4$ & 5.20 & 2 & 4 & 52.1 & 7.25 $\times 10^2$ \\ 
2.6 $\times 10^4$ & 5.19 & 1 & 4 & 87.8 & 9.90 $\times 10^2$ \\ 
3.8 $\times 10^4$ & 5.18 & 1 & 3 & 71.5 & 9.10 $\times 10^3$ \\ 
5.4 $\times 10^4$ & 5.17 & 1 & 3 & 72.5 & 1.38 $\times 10^4$ \\ 
7.8 $\times 10^4$ & 5.15 & 2 & 2 & 35.8 & 2.77 $\times 10^4$ \\ 
1.0 $\times 10^5$ & 5.13 & 1 & 2 & 60.4 & 5.15 $\times 10^4$ \\ 
1.3 $\times 10^5$ & 5.11 & 1 & 2 & 61.2 & 3.17 $\times 10^5$ \\ 
1.5 $\times 10^5$ & 5.09 & 1 & 1 & 43.9 & 4.82 $\times 10^7$ \\   
\bottomrule
\end{tabular}
\end{center}
\label{table:period10}
\end{table}

\section{{Discussion and Summary}}
We have analyzed fifteen different pre-white dwarf model sequences which have various amounts of hydrogen.
In all the sequences, we have found the pulsational instabilities due to the $\varepsilon$-mechanism.  
It is remarkable that even a pre-DA white dwarf model sequence with the poorest amount of hydrogen  has unstable modes excited by the hydrogen burning. 
The instability region in an evolution diagram, such as the $\log T_{\rm eff}$-$\log g$ diagram or the H-R diagram, is {very} sensitive to the amount of hydrogen. 
Thus, it {should be} possible to constrain observationally the thickness of the envelope in pre-DA white dwarfs from {the presence of oscillations} with periods of 40-200\,s in {these hot white dwarfs}. 

\citet{Handler1998} searched for periodic light variations among low-mass hot DA stars exhibiting helium in their optical spectra (classified as the DAO stars), aiming at discovering low-order g-modes which have theoretically been predicted by \citet{Charpinet1997} to be excited by the $\varepsilon$-mechanism, but no pulsations with amplitudes higher than $\sim\!1$\,mmag were detected.  
{However, optical spectra of these DAO stars were recently reanalyzed along with other DAO stars and it turned out that none of them is a post-EHB star \citep{Gianninas2010}. Rather, they are now considered to be post-AGB stars, as shown in figure\,\ref{fig:6}.
Hence these stars are relevant to testing the predictions given in the present paper. 
Absence of observed pulsation in those stars would imply that either (i) the pulsation amplitudes are smaller than the detection limit, (ii) our current post-AGB evolution models do not match well the structures of the real pre-white dwarfs with hydrogen-dominated atmospheres, or (iii) the atmospheric parameters deduced from spectroscopic observations are significantly different from the real ones. Any of the case (i), (ii) and (iii) makes a big impact on our understanding of physics on white dwarfs. Thus, further observational tests are highly required.}

All the known pulsations of white dwarfs are excited in the ionization zone of a certain chemical element
({hydrogen, helium, and other ions}), which is located near the photosphere \citep{Fontaine2008, Winget2008}. 
In contrast, the pulsations analyzed in this paper are excited mostly at the bottom of the hydrogen envelope of pre-white dwarfs. 
{This implies that} the oscillations of such pre-white dwarfs may allow us to investigate {deeper regions}. 
This opens a new window {for} asteroseismology to unveil the invisible interior of pre-white dwarfs and {test} the relevant unsolved physics.

\bigskip
\section*{Acknowledgments}
We wish to thank M. Takata, M. Godart, O. Benomar, T. Sonoi, and T. Sekii for their useful comments.

\end{document}